\documentclass[twocolumn,superscriptaddress,preprintnumbers,amsmath,amssymb,longbibliography]{revtex4-1}
\usepackage{graphicx}
\usepackage{epstopdf}
\usepackage{dcolumn}
\usepackage{bm}
\usepackage{color}
\usepackage{braket}
\usepackage{dcolumn}
\usepackage{mathrsfs}
\usepackage[colorlinks,linkcolor=blue,anchorcolor=blue,citecolor=blue,urlcolor=blue]{hyperref}

\makeatletter
\@ifundefined{textcolor}{}
{%
 \definecolor{BLACK}{gray}{0}
 \definecolor{WHITE}{gray}{1}
 \definecolor{RED}{rgb}{1,0,0}
 \definecolor{GREEN}{rgb}{0,1,0}
 \definecolor{BLUE}{rgb}{0,0,1}
 \definecolor{CYAN}{cmyk}{1,0,0,0}
 \definecolor{MAGENTA}{cmyk}{0,1,0,0}
 \definecolor{YELLOW}{cmyk}{0,0,1,0}
}

\begin{document}
\UseRawInputEncoding

\title{Nonclassical correlated deterministic single-photon pairs for a trapped atom in bimodal cavities}

\author{Zhong Peng}
\affiliation{Guangdong Provincial Key Laboratory of Quantum Metrology and Sensing $\&$ School of Physics and Astronomy, Sun Yat-Sen University (Zhuhai Campus), Zhuhai 519082, China}

\author{Yuangang Deng}
\email{dengyg3@mail.sysu.edu.cn}
\affiliation{Guangdong Provincial Key Laboratory of Quantum Metrology and Sensing $\&$ School of Physics and Astronomy, Sun Yat-Sen University (Zhuhai Campus), Zhuhai 519082, China}

\date{\today}

\begin{abstract}
Single photons and single-photon pairs, inherently nonclassical in their nature, are fundamental elements of quantum sciences and technologies. Here, we propose to realize the nonclassical correlated deterministic photon pairs at the single-photon level for a single atom trapped in bimodal cavities. It is shown that the photon emissions for bimodal cavities exhibit a high single photon purity and relatively large intracavity photon numbers, which are ascribed to the combination of optical Stark shift enhanced energy-spectrum anharmonicity and quantum interference suppressed two-photon excitation. Our scheme generates photon-photon pairs with strong photon blockade for cavity modes via constructing a highly tunable cavity-enhanced deterministic parametric down-conversion process. Furthermore, we show that the cross-correlation function between the two cavity modes vastly violates the Cauchy-Schwarz inequality of the classical boundary, which unambiguously demonstrates the nonclassicality of the single-photon pairs. Our result reveals a prominent strategy to generate the high-quality two-mode single-photon sources with strong nonclassical correlation, which could provide versatile applications in distribution of quantum networks, long distance quantum communication, and fundamental tests of quantum physics.
\end{abstract}

\maketitle
\section{Introduction}

The generation of high-quality deterministic single-photon sources is a crucial building-block for quantum information science, quantum computing and quantum cryptography~\cite{Northup2014,awschalom2018quantum,Arakawa2020,blais2020quantum}.
A paradigmatic mechanism for generation of single photons is utilizing the photon blockade (PB) \cite{PhysRevA.41.475,PhysRevLett.55.2790,Deutsch1997Kerr}, in which the quantum
statistics of the photon emissions exhibit a sub-Poissonian distribution since one photon excitation will block a second photon excitation. Up to now, the controllable single-photon states have been extensively studied by using the mechanisms of conventional PB and unconventional PB. The former mechanism requires a strong optical nonlinearity with large enough energy-spectrum anharmonicity in the systems, ranging from cavity quantum electrodynamics (QED)~\cite{McKeever2004,Kimble2005pb,bajcsy2013photon,Radulaski2017,Hamsen2017,zhang2018high,Trivedi2019,Tang2019,tang2021tunable} to circuit QED~\cite{Lang2011,Hoffman2011,Liu2014,PhysRevA.97.013851,wang2018photon}, Kerr-type nonlinear systems~\cite{Deutsch1997Kerr,Liao2010,Ferretti2010,Miranowicz2013,li2017cascaded,PhysRevLett.121.153601,Ghosh2019,lin2020kerr}, and optomechanical resonators~\cite{Rabl2011,Stannigel2012,Liao2013optomechanical,Ng2016,Zhu2018,PhysRevA.99.043818,xu2020quantum,Wang_2020}. The latter mechanism relies on the destructive quantum interference by constructing different quantum transition pathways~\cite{PhysRevLett.104.183601,PhysRevLett.108.183601,tang2015quantum,Snijders2018,vaneph2018,li2019nonreciprocal,zubizarreta2020}. Thus the two-photon excitation can be completely eliminated even with a weak atom-cavity coupling. In addition to these significant advances, the strong PB by utilizing the optical Stark-shift enhanced energy-spectrum anharmonicity has been proposed for Jaynes-Cummings model and electromagnetically induced transparency in Refs.~\cite{Tang2019,tang2021tunable}.

Meanwhile, the availability of nonclassical correlated photon pairs could offer new resources for exploration of intriguing quantum physics~\cite{RevModPhys.84.777,RevModPhys.86.419,RevModPhys.90.025004}. The spontaneous parametric down-conversion (SPDC) of nonlinear optical processes~\cite{Donald1970Parametric,ou1999cavity}, as a paradigmatic mechanism for realizing nonclassical correlated photon pairs, has been studied in numerous quantum applications, such as quantum cryptography~\cite{naik2000entangled,lasota2020optimal}, entangled photons~\cite{kwiat1995new,wang2004polarization,zhang2011preparation,clausen2014source}, quantum imaging~\cite{lemos2014imaging}, quantum memories~\cite{heshami2016quantum,seri2017quantum,tsai2020quantum} and quantum networks~\cite{kimble2008quantum,hu2021long}. The aforementioned schemes based on SPDC are probabilistic since the generated photon pairs are random, albeit the recent advances for photon-pair sources have been proposed by employing deterministic down-conversion~\cite{PhysRevLett.117.203602,PhysRevA.94.053814}. By contrast, a deterministic photon source can be achieved in the isolated single atom cavity system~\cite{McKeever2004}. Unfortunately, the deterministic source reduces to a probabilistic one with including the inevitable losses for both atom and cavity, corresponding to nonzero multiphoton excitations. In addition, due to the intrinsic weak nonlinear interactions between photon pairs, the experimental realization of high-purity photon pairs at high rates remains a challenge. Therefore, the realization of deterministic photon-pair sources in single-photon level beyond strong coupling regime will provide unprecedented opportunities in quantum-physics community and motivate numerous implementations in photonic quantum information processing~\cite{RevModPhys.87.347,flamini2018photonic,slussarenko2019photonic} and many-body physics~\cite{chang2014quantum,chang2008crystallization,tangpanitanon2020many}.

\begin{figure}[htb]
\includegraphics[width=0.9\columnwidth]{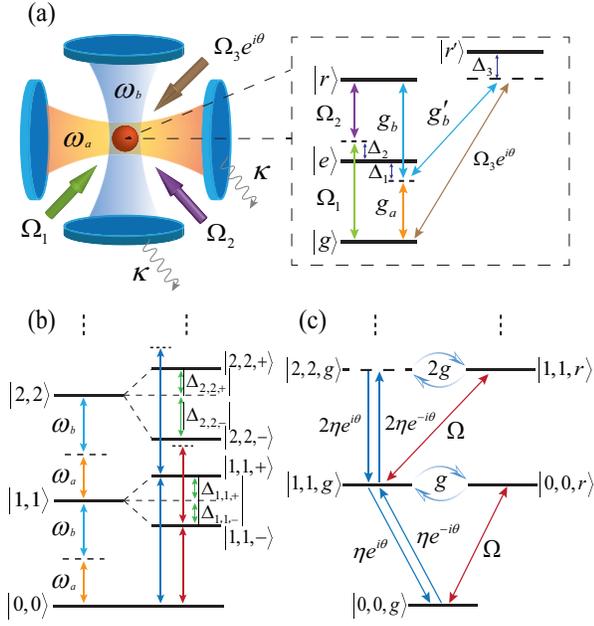}
\caption{(a) Schematic of the system for realizing nonclassical correlated deterministic single-photon pairs in bimodal cavities coupled to a single atom. (b) The typical anharmonic energy spectrum with ignoring the weak atomic pump field and effective cavity-driven field. (c) Quantum transition paths based on the Hamiltonian  (\ref{Hamiltonian2}). The strong PB is generated with eliminating the occupation probability of two-photon excitation state $|2,2,g\rangle$ (dashed line) by using destructive quantum interference.} \label{Fig1}
\end{figure}

In this paper, we propose to generate nonclassical correlated deterministic single-photon pairs for a single atom trapped in bimodal cavities. Here the deterministic cavity-enhanced  parametric down-conversion Hamiltonian is realized with a high tunability. It is shown that the interplay of optical Stark-shift enhanced energy-spectrum anharmonicity and quantum interference suppressed two-photon excitation gives rise to strong PB for both the two cavity modes. In particular, the second-order correlation function with the optimal phase and driven amplitude of cavities can be low to $10^{-4}$ at a moderate Stark shift, corresponding to a relatively large intracavity photon numbers as well. Remarkably, the photon emissions for the two-mode cavities exhibit a strong nonclassical correlation with violating Cauchy-Schwarz inequality of classical boundary, which host a high single photon purity with identical photon numbers. Our proposal will be valuable for realizing deterministic high-quality single-photon sources and nonclassical correlated single-photon pairs sources.

This paper is organized as follows. In Sec.~\ref{model}, we introduce the model and Hamiltonian for realizing a deterministic parametric down-conversion process. Section~\ref{stark} is devoted to derive the Stark-shift enhanced energy-spectrum anharmonicity. In Sec.~\ref{PBQI}, we present strong PB with quantum interference. In Sec.~\ref{pairs}, we study nonclassical correlated single-photon pairs of the single atom-cavity system. Finally, we conclude in Sec.~\ref{Con}.

\section{Model and Hamiltonian}
\label{model}
Without loss of generality, we consider the high-finesse bimodal cavities coupled to a single four-level atom, where $\omega_a$ ($\omega_b$) is the bare frequency for the single-mode optical cavity A (B). Figure~\ref{Fig1}(a) shows the level structure and laser configuration of the single atom-cavity system. The four relevant energy levels include one electronic ground state $\ket{g}$ and three excited states $\ket{e}$, $\ket{r}$ and $\ket{r'}$, where $\ket{r}$ is an elaborately long-lived electronic orbital state with the hyperfine splitting $\hbar \Delta_{3}$ between the states $\ket{r}$ and $\ket{r'}$. As can be seen, the atomic transition $\ket{g}$ $\leftrightarrow$ $\ket{e}$ ($\ket{e}$ $\leftrightarrow$ $\ket{r}$) is coupled to the cavity mode A (B) with single atom-cavity coupling strength $g_a$ ($g_b$) and atom-cavity detuning $\Delta_1$. Moreover, the atomic transition $\ket{e}$ $\leftrightarrow$ $\ket{r'}$ is far-off resonance driven by the cavity field B with single atom-cavity coupling $g'_b$ and large atom-cavity detuning $\Delta_3$.

In our laser configuration, the single atom is also illuminated by a pair of classical pump fields with Rabi frequencies $\Omega_1$ and $\Omega_2$, corresponding to the atom-pump detuning $\Delta_2$. To generate an effective driven field of bimodal cavities, we consider applying a strong pump field to drive the atomic transition $\ket{g}$ $\leftrightarrow$ $\ket{r'}$ with Rabi frequency $\Omega_3  e^{i\theta}$ as depicted in Fig.~\ref{Fig1}(a). We should emphasize that the tunable phase $\theta$ is gauge invariant and will play an essential role in realizing strong PB for bimodal cavities by employing the quantum interference.

In the far dispersive regime, $|\Delta_i|\gg \{g_a,g_b,\Omega_{i=1,2,3}\}$, the far-off resonance atomic excited states $\ket{e}$ and $\ket{r'}$ can be adiabatically eliminated. Under the rotating-wave approximation, the relevant time-independent Hamiltonian for the single atom-cavity system reads (see
Appendix A for details)
\begin{align}\label{Hamiltonian2}
\hat{\cal H}/\hbar &= \Delta_{a}\hat{a}^{\dagger}\hat{a} + \Delta_{b}\hat{b}^{\dagger}\hat{b} + (\Delta_{r} - U_{a}\hat{a}^{\dagger}\hat{a} + U_{b}\hat{b}^{\dagger}\hat{b})\hat{\sigma}_{rr} \nonumber \\
&+ [(g\hat{a}^{\dagger}\hat{b}^{\dagger}+\Omega)\hat{\sigma}_{gr} + \eta e^{-i\theta}\hat{a}^{\dagger}\hat{b}^{\dagger}\hat{\sigma}_{gg} + {\rm H.c.}],
\end{align}
where $\hat{a}$ $(\hat{b})$ is the operator of cavity mode A (B) with cavity-light detuning $\Delta_a$ ($\Delta_b$), $\Delta_r$ is the effective two-photon detuning, and $U_a=-g^{2}_{a}/\Delta_{1}$ ($U_b=-g^{2}_{b}/\Delta_{1}$) is the optical Stark shift for cavity mode A (B). $\hat{\sigma}_{kl}=\ket{k}\bra{l}(k,l=g,r)$ denotes the atomic transition operator, $g=-g_{a}g_{b}/\Delta_{1}$ ($\Omega=-\Omega_{1}\Omega_{2}/\Delta_{1}$) is cavity (pump) field coupling strength, and $\eta = \Omega_{3}g_{a}g^{\prime}_{b}/\Delta_{1}\Delta_{3}$ is the weak cavity-driven amplitude. Here, we neglect terms with the high-frequency prefactor $e^{\pm i(\Delta_1-\Delta_2)t}$ in Eq.~(\ref{Hamiltonian2}) when $|\Delta_1-\Delta_2|\gg \{g_a,g_b,\Omega_{i=1,2}\}$.

Remarkably, the Hamiltonian (\ref{Hamiltonian2}) essentially describes a two-photon Raman-assisted deterministic parametric down-conversion process, where the photon excitations for the bimodal cavities are simultaneously created or destroyed. As we will see below, high-quality single-photon pairs are expected, corresponding to a strong nonclassical correlation of the single photons emitted by the separated bimodal cavities simultaneously. In contrast to the generation of probabilistic photon pair sources with SPDC~\cite{fasel2004high,PhysRevLett.106.013603,caspani2017integrated,senellart2017high,puigibert2017heralded}, the realized nonclassicality two-mode photon pairs can be ascribed to the deterministic sources in our isolated single quantum system. In the presence of atom and cavities dissipations, the multiphoton excitations can be completely suppressed by employing the optical Stark-shift enhanced energy spectrum anharmonicity and quantum interference suppressed two-photon excitation.

To further investigate the quantum statistics and correlations of photon emissions, we take into account the dissipations of the atomic field and bimodal cavities by numerically solving the master equation in the steady state using the quantum optics toolbox~\cite{Tan_1999}. As a result, the time evolution of the density matrix $\rho$ for the cavities and atomic field obeys the master equation, ${d\rho}/{dt}=\mathcal{L} \rho$, corresponding the Liouvillian superoperator defined as
\begin{align}\label{master equation}
\mathcal{L} \rho&= -i[{\cal {\hat{H}}},\rho] + \sum_{\hat{o}}\frac{\kappa_o}{2} \mathcal{D} [\hat{o}] \rho + \frac{\gamma}{2} \mathcal{D} [\hat{\sigma}_{gr}] \rho + \frac{\gamma_d}{2} \mathcal{D} [\hat{\sigma}_{rr}] \rho,
\end{align}
where $\kappa_o$ is the decay rate of cavity field with ${o}={a}$ and ${b}$, $\gamma$ is the atomic spontaneous emission rate, $\gamma_d$ is the pure dephasing factor of the atomic field, and $\mathcal{D} [\hat{o}] \rho = 2\hat{o} \rho \hat{o}^{\dagger} - \hat{o}^{\dagger} \hat{o} \rho - \rho \hat{o}^{\dagger} \hat{o}$ is the Lindblad type of dissipation. In the steady-state solution, $\mathcal{L} \rho_s=0$, the intracavity photon number for cavity mode $\hat{o}$ satisfies, $n^{(o)}_s = {\rm Tr}[\hat{o}^{\dagger}\hat{o}\rho_s]$. In general, the second order correlation function, which is a key physical quantity for characterizing  PB and nonclassical correlation, is defined as~\cite{PhysRev.130.2529}
\begin{align}
g_{oo'}^{(2)}(\tau)=\frac{{\rm Tr}[\hat{o}^\dagger(t)\hat{o'}^\dagger(t+\tau)
\hat{o'}(t+\tau)\hat{o}(t)\rho_s]}{{\rm Tr}[\hat{o}^\dagger(t)
\hat{o}(t)\rho_s]\times {\rm Tr}[\hat{o'}^\dagger(t)
\hat{o'}(t)\rho_s]}.
\end{align}
For the zero time interval ($\tau=0$), $g_{oo'}^{(2)}(\tau)$ is directly reduced to the equal time second-order auto-correlation function $g_{aa}^{(2)}(0)$ [$g_{bb}^{(2)}(0)$] for the optical mode A (B) with $o=o'$, whose value is used to characterize the purity of single-photon emissions~\cite{senellart2017high}. As to $o\neq o'$, $g_{ab}^{(2)}(0) = g_{ba}^{(2)}(0)$ represents the cross-correlation function between separated bimodal cavities. Furthermore, the time-dependent second-order correlation function $g_{oo'}^{(2)}(\tau)$ can be straightforwardly calculated using the quantum regression theorem~\cite{carmichael1999statistical}. Therefore, the quantum statistics for PB should satisfy two conditions: $g_{oo}^{(2)}(0)<1$ to ensure the sub-Poissonian statistics and $g_{oo}^{(2)}(0)<g_{oo}^{(2)}(\tau)$ to ensure the photon antibunching. Moreover, the second-order correlation functions can be experimentally measured via Hanbury, Brown, and Twiss interferometer~\cite{Kimble2005pb}.

As to the experimental feasibility, our proposal could be applicable to alkaline-earth-metal atoms~\cite{PhysRevLett.118.263601,PhysRevLett.117.220401,kolkowitz2017spin,bromley2018dynamics} and Rydberg atoms~\cite{henkel2010three,madjarov2020high} by employing the advantages of energy-level structures. In our numerical simulation, we assume the bimodal cavities have the equal decay rates with $\kappa=\kappa_a=\kappa_b$ for simplicity. The cavity decay rate $\kappa=2\pi\times800$ kHz~\cite{leonard2017supersolid,leonard2017monitoring} is fixed as the energy scale of the system, then the corresponding atomic decay rate for the long-lived excited state $\gamma/\kappa=0.01$ and pure dephasing $\gamma_d/\kappa=0.01$. We take the single atom-cavity coupling $g/\kappa=2$, the weak atomic pump field $\Omega/\kappa=0.1$, and $\Delta_{a} = \Delta_{b} = \Delta_{r}/2$. Thus the tunable parameters in our atom-cavity system are reduced to cavity-light detuning $\Delta_a$ and optical Stark shift $U_a$ with satisfying $U_aU_b=g^2$. Moreover, the effective cavity-driven strength $\eta$ and its phase $\theta$ are self-consistently determined by applying the quantum interference conditions in our numerical simulation.

\section{Stark shift enhanced PB}
\label{stark}
For first ignoring the weak driving field of the bimodal cavities and atomic classical pump field, the analytical energy spectrum of the system is obtained by diagonalizing Hamiltonian (\ref{Hamiltonian2}). We note that the difference of the excitation number between cavity mode A and B is $zero$ due to the deterministic parametric down-conversion process. Therefore the restricted Hilbert spaces for the single atom-cavity system are $\ket{n-1,n-1,r}$ and $\ket{n,n,g}$, where $n$ denotes the photon number excitation of single-cavity mode. Explicitly, the energy eigenvalues of system are given by
\begin{align}\label{Energy spectrum}
E_{n,n,\pm} &= 2 n \Delta_{a} - \frac{1}{2}(n - 1)U_0 \pm \frac{1}{2}\sqrt{(n - 1)^{2}U_0^2 + 4g^{2}n^{2}},
\end{align}
corresponding to the energy splittings for the $n$th pair of dressed states $\ket{n,n,\pm}$
\begin{align}\label{splitting}
\Delta_{n,n,\pm} &= \pm \frac{1}{2}\sqrt{(n - 1)^{2}U_0^2 + 4g^{2}n^{2}}  - \frac{1}{2}(n - 1)U_0,
\end{align}
where $+$ ($-$) represents the higher (lower) branch and $U_{0} = U_a - U_b$ is Stark-shift difference of bimodal cavities. It is clear that the vacuum Rabi splitting $\Delta_{1,1,\pm}= \pm g$ is independent on the value of $U_0$, which is contrast to the realization of strong PB proposed in Ref.~\cite{Tang2019} by utilizing the Stark shift to enhance the vacuum Rabi splitting.
However, the anharmonic ladder of dressed states for the multiphoton excitations ($n\geq 2$) is enhanced by the large value of $U_0$. In Fig.~\ref{Fig1}(b), we show the typical anharmonicity ladder of the energy spectrum with $\eta=0$ and  $\Omega=0$. We emphasize that the energy spectrum is immune to the weak atomic pump and cavities driven fields, i.e., $|\Omega/g| \ll 1$ and $|\eta/g| \ll 1$. As expected, the strong energy-spectrum anharmonicity indicates that the strong PB for bimodal cavities emissions can be realized.

\begin{figure}[htb]
\includegraphics[width=0.9\columnwidth]{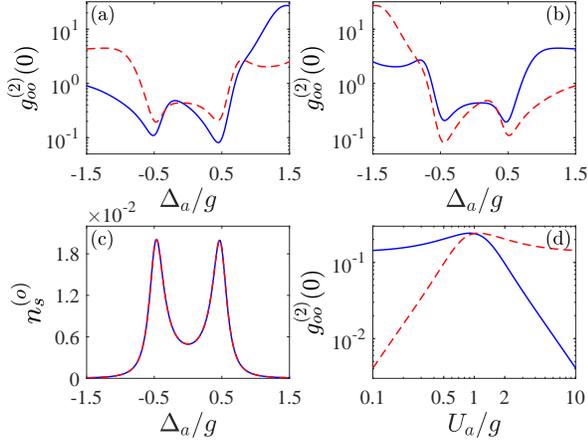}
\caption{ The second-order correlation function $g_{oo}^{(2)}(0)$ as a function of $\Delta_a$ for (a) $U_{a}/g=2$  and (b) $U_{a}/g=1/2$, respectively. (c) The steady-state photon number $n_s^{(o)}$ as a function of $\Delta_a$ for $U_{a}/g=2$. Here, the solid blue line (dashed red line) denotes the result of cavity mode A (B). (d) The $U_{a}/g$ dependence of $g_{oo}^{(2)}(0)$ for cavity mode A at the red sideband with $\Delta_a/g = 1/2$ (solid blue line) and for cavity mode B at the blue sideband with $\Delta_a/g = -1/2$ (dashed red line), respectively. In (a)--(d), the other parameter is $\eta/g=0$.} \label{Fig2}
\end{figure}

To investigate the PB enhanced by optical Stark shift, we calculate the photon quantum statistics for fixing $\eta/g=0$, where the quantum interference conditions are invalid when the effective cavity-driven field disappears~\cite{tang2015quantum,Tang2019}. Figures.~\ref{Fig2}(a) and \ref{Fig2}(b) display the equal time second-order correlation function as a function of $\Delta_a$ for $U_{a}/g=2$ and $U_{a}/g=1/2$, respectively. For $U_a/g\neq 1$, it is shown that the line shapes of $g_{aa}^{(2)}(0)$ (solid blue line) and $g_{bb}^{(2)}(0)$ (dashed red line) exhibit the red-blue asymmetric behavior with respect to cavity-light detuning $\Delta_a$. Interestingly, the steady-state photon numbers $n_s^{(a)}$ and $n_s^{(b)}$ for the bimodal cavities with the same dissipations always coincide exactly due to the parametric down-conversion process, as displayed in Fig.~\ref{Fig2}(c). Moreover, it is clear that $n_s^{(o)}$ exhibits a red-blue symmetric profile and possesses two peaks at single-photon resonance with $\Delta_a/g=\pm 1/2$. The reason is that the energy eigenvalues $E_{1,1,\pm}$ in Eq.~(\ref{Energy spectrum}) is independent of the value of Stark shift $U_0$.

As to the photon quantum statistics, we find that the minimum value of $g_{aa}^{(2)}(0)$ [$g_{bb}^{(2)}(0)$] for cavity mode A (B) occurs at the red-sideband (blue-sideband) when $U_a/g >1$ ($U_a/g <1 $). Figure.~\ref{Fig2}(d) displays the result of $g_{oo}^{(2)}(0)$ versus $U_{a}/g$. We find that the second-order correlation function for the cavity mode A at the red sideband is equal to the cavity mode B at the blue sideband when $U_a/g=1$ yielding $U_a/U_b=1$. For $U_a/g>1$, the antibunching amplitude for cavity mode A (solid blue line) is rapidly enhancing with increasing $U_a$. As to the regime of $U_a/g<1$ ($U_b/g>1$), it is clear that $g_{bb}^{(2)}(0)$ for cavity mode B (dashed red line) is monotonically decreasing with decreasing $U_a$. This result can be readily understood through the calculation of the energy spectrum dominated by the Hamiltonian (\ref{Energy spectrum}), where the large Stark shift gives rise to the large energy-spectrum anharmonicity. As can be seen, the strong PB for cavity emission, which is defined as $g_{oo}^{(2)}(0)<0.01$ requires a large Stark shift with $U_a/g\gg 1$ or $U_a/g\ll 1$ ($U_b/g\gg 1$).

\section{Strong PB with quantum interference}
\label{PBQI}
For the conventional PB in cavity QED~\cite{McKeever2004,Kimble2005pb,bajcsy2013photon,Radulaski2017,Hamsen2017,zhang2018high,Trivedi2019,Tang2019,tang2021tunable} , the strong energy spectrum anharmonicity is required in the strong single atom-cavity coupling limit. By contrast, the unconventional PB can be achieved by utilizing quantum interference beyond the strong coupling regime. By employing the destructive quantum interference, the two-photon excitations can be completely suppressed, which has been proposed in Jaynes-Cummings model~\cite{tang2015quantum,Tang2019}. Analogously, the present deterministic parametric down-conversion Hamiltonian (\ref{Hamiltonian2}) for bimodal cavities should also exist the quantum interference mechanism with combination of classical atomic pump field and effective cavity driven field. To obtain the quantum interference conditions, the wave function of the system is formally given as
\begin{align}
\ket{\psi} = \sum_{n=0}^{\infty}C_{n,n,g}\ket{n,n,g} + \sum_{n=0}^{\infty}C_{n,n,r}\ket{n,n,r},
\end{align}
where $|C_{n,n,g}|^{2}$ and $|C_{n,n,r}|^{2}$ denote the atomic occupation probability of the system for the eigenstate $\ket{n,n,g}$ and $\ket{n,n,r}$, respectively. Without loss of generality, we can safely neglect the multiphoton excitations ($n>2$) in PB regime. Thus, the wave function of the system with truncating the two-photon excitations subspaces reads
\begin{align}
\ket{\psi} = \sum_{n=0}^{2}C_{n,n,g}\ket{n,n,g} + \sum_{n=0}^{1}C_{n,n,r}\ket{n,n,r}.
\end{align}
Substituting the wave function $\ket{\psi}$ into the Schr\"{o}dinger equation, we can derive a set of evolution equations
\begin{align}\label{interference}
i\dot{C}_{0,0,g} &= \Omega C_{0,0,r} + \eta e^{i\theta}C_{1,1,g}, \nonumber \\
i\dot{C}_{1,1,g} &= \tilde{\Delta}_{a}C_{1,1,g} + gC_{0,0,r} + \Omega C_{1,1,r} + \eta e^{-i\theta}C_{0,0,g}  \nonumber \\
&+ 2\eta e^{i\theta}C_{2,2,g}, \nonumber \\
i\dot{C}_{0,0,r} &= \tilde{\Delta}_{r}C_{0,0,r} + gC_{1,1,g} + \Omega C_{0,0,g}, \nonumber \\
i\dot{C}_{2,2,g} &= 2\tilde{\Delta}_{a}C_{2,2,g} + 2gC_{1,1,r} + 2\eta e^{-i\theta}C_{1,1,g}, \nonumber \\
i\dot{C}_{1,1,r} &= (\tilde{\Delta}_{a} + \tilde{\Delta}_{r} - U_{0})C_{1,1,r} + 2gC_{2,2,g}  + \Omega C_{1,1,g},
\end{align}
with $\tilde{\Delta}_{a} = 2\Delta_{a} - i2\kappa$ and $\tilde{\Delta}_{r} = 2\Delta_{a} - i\gamma$ for shorthand notations. In the steady-state solutions, the two-photon excitation can be completely eliminated with $C_{2,2,g} = 0$, which indicates that the single-photon emission is obtained. As a result, the nontrivial solutions of Eq.~(\ref{interference}) are given by
\begin{align}\label{QI}
&\theta_{\rm{opt}} = -\arctan\left(\frac{2\kappa + \gamma}{4\Delta_a - U_{0}}\right),       \nonumber \\
&\eta_{\rm{opt}} = {g\Omega}/{{\cal R}},
\end{align}
with ${\cal R}=\sqrt{(4\Delta_a-U_{0})^2 + (2\kappa + \gamma)^2}$.

Remarkably, the optimal parameters $\theta_{\rm{opt}}$ and $\eta_{\rm{opt}}$ denote the quantum interference conditions for realization of strong PB. In contrast to the optimal atomic pump field in Ref.~\cite{Tang2019,tang2015quantum}, the quantum interference conditions in our system can be highly tuned by the effective cavity-driven field. For fixing the single-photon resonance, i.e., $\Delta_{a} = \pm g/2$, the strong PB with a large cavity output is naturally expected by using the obtained optimal cavity-driven amplitude $\eta_{\rm{opt}}$ and its phase $\theta_{\rm{opt}}$ in Eq.~(\ref{QI}).

\begin{figure}[htb]
\includegraphics[width=0.9\columnwidth]{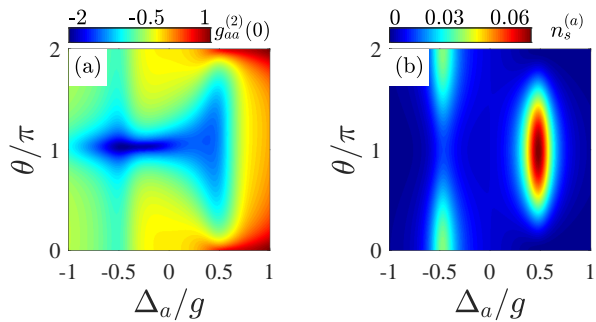}
\caption{ (a) $g_{aa}^{(2)}(0)$ and (b) $n^{(a)}_s$ as a function of $\Delta_a$ and $\theta$ with $U_a/g=2$. Here, the cavity-driven amplitude $\eta$ is self-consistently calculated by using the quantum interference conditions in Eq.~(\ref{QI}). The colors with blue-red gradient shading denote the values of $\log_{10}[g_{aa}^{(2)}(0)]$ in (a) and $n^{(a)}_s$ in (b), respectively.} \label{Fig3}
\end{figure}

To process further, the essential mechanism for destructive quantum interference can be well understood by the quantum transition paths of the atom-cavity system, as shown in Fig.~\ref{Fig1}(c). There exist two quantum transition pathways from single-photon excitation to two-photon excitation. Explicitly, the transition $\left|1,1,g\right\rangle \xrightarrow{2\eta e^{-i\theta}} \left|2,2,g\right\rangle$ is induced by the effective driven field of cavities. Meanwhile, the transition $\left|1,1,g\right\rangle \xrightarrow{\Omega} \left|1,1,r\right\rangle \xrightarrow{2g} \left|2,2,g\right\rangle$ is related to the atomic pump field and cavity fields. Actually, the optical Stark-shift induced effective cavity-driven term in Hamiltonian (\ref{Hamiltonian2}) plays an important role in building the quantum interference. It is clear that the quantum interference conditions in Eq.~(\ref{QI}) will vanish when $\eta=0$. More importantly, the forbidden transition of $\left|0,0,r\right\rangle \rightarrow \left|1,1,r\right\rangle$ will also be beneficial to suppress the two-photon excitation in our laser configuration, which is in contrast to the previous proposals~\cite{Tang2019,tang2015quantum} that the single-mode cavity is directly driven by the external laser field.

We should emphasize that the quantum interference mechanism for generating strong PB also requires a sufficient energy-spectrum anharmonicity. In fact, the multiphoton ($n\geq 3$) excitations could not be well suppressed in the weak atom-cavity coupling regime. We check that the second-order correlation function $g_{aa}^{(2)}(0)$ and $g_{bb}^{(2)}(0)$ is immune to the weak optical Stark shift when $U_a/g\ll1$ and $U_b/g\ll 1$, respectively. Without loss of generality, we will study the photon quantum statistics for cavity mode A with focusing on $U_a/g>1$.

Figures~\ref{Fig3}(a) and \ref{Fig3}(b) show the numerical results of $g_{aa}^{(2)}(0)$ and $n^{(a)}_s$ in the $\Delta_a$-$\theta$ parameter plane with quantum interference conditions in Eq.~(\ref{QI}) for $g/\kappa=2$ and $U_a/g=2$. Compared with the results of $\eta=0$ (without quantum interference), both $g_{aa}^{(2)}(0)$ and $n^{(a)}_s$ for nonzero $\eta$ exhibit obviously asymmetric behaviors, which depend on the sign of cavity-light detuning $\Delta_a$. In particular, the strong PB with $g_{aa}^{(2)}(0)<0.01$ is realized near the blue sideband beyond the strong coupling limit. And the sufficiently large cavity emission $n^{(a)}_s$ can also be obtained in a large parameter region with strong PB [Fig.~\ref{Fig3}(b)]. Remarkably, the strong PB existing in a large parameter regime will facilitate the experimental feasibility even the system parameters deviating from quantum interference conditions.

\begin{figure}[htb]
\includegraphics[width=0.9\columnwidth]{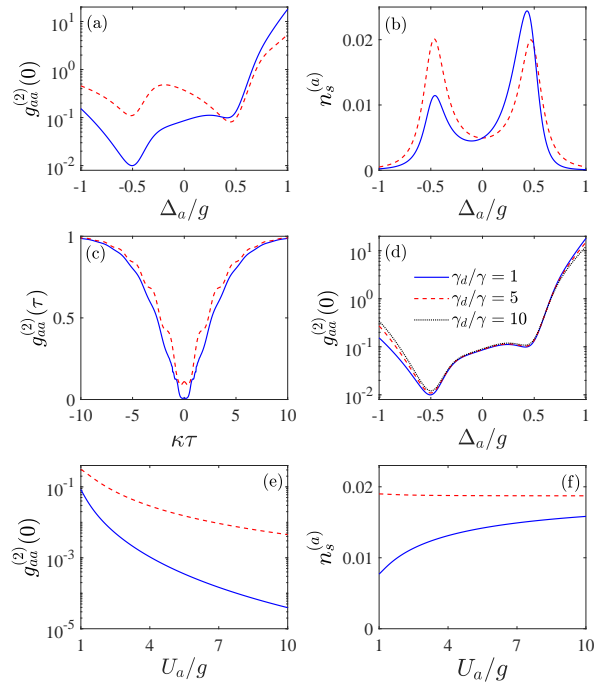}
\caption{$\Delta_a$ dependance of (a) $g_{aa}^{(2)}(0)$ and (b) $n^{(a)}_s$. (c) Time interval $\tau$ dependence of $g^{(2)}_{aa}(\tau)$ for $\Delta_{a}/g=-1/2$. (d) $\Delta_a$ dependence of $g_{aa}^{(2)}(0)$ for different values of pure dephasing $\gamma_d$. In (a)--(d), we fix the Stark shift $U_a/g=2$. (e) $g_{aa}^{(2)}(0)$  and (f) $n^{(a)}_s$  versus $U_{a}/g$ for fixing the blue sideband with $\Delta_{a}/g=-1/2$. Excluding (d), the solid blue line (dashed red line) represents the result with (without) quantum interference conditions of Eq.~(\ref{QI}).} \label{Fig4}
\end{figure}

Furthermore, Figs.~\ref{Fig4}(a) and \ref{Fig4}(b) show $g_{aa}^{(2)}(0)$ and $n^{(a)}_s$ as a function of $\Delta_a$ with (without) quantum interference conditions. Although the steady-state photon number at blue sideband is decreased, the second-order correlation function $g_{aa}^{(2)}(0)$ of $\eta\neq 0$ (solid blue line) can be reduced more than one orders of magnitude compared with the results of $\eta=0$ (dashed red line) even at a moderate Stark shift, i.e., $U_a/g=2$. As to red sideband $\Delta_a/g=1/2$, we find that cavity emission of $n^{(a)}_s$ is obviously growing with $\eta\neq 0$, albeit $g_{aa}^{(2)}(0)$  is roughly unchanged.

In Fig. \ref{Fig4}(c), we further calculate the time interval $\tau$ dependence of the second-order correlation function $g^{(2)}_{aa}(\tau)$. It is clear that the transmitted photons exhibit the strong PB, with $g^{(2)}_{aa}(0)=0.01$ and photon antibunching with $g^{(2)}_{aa}(0)<g^{(2)}_{aa}(\tau)$. Moreover, the lifetime of the PB is significantly longer than the typical value of $\tau \sim 1/\kappa$ since $\gamma/\kappa\ll 1$ in our system. As to photon quantum statistics, we check that $g^{(2)}_{aa}(0)$ with respect to $\Delta_a$ is immune to the weak atomic pure dephasing when $\gamma_d/\gamma\leq 10$, as displayed in Fig. \ref{Fig4}(d). In particular, we remark that the qualitative results for both $g_{aa}^{(2)}(0)$ and $n_s^{(a)}$ are robust against the variations of parameters $\kappa_a/\kappa_b$, albeit the values of them depend on the specific choice of $\kappa_a$ and $\kappa_b$ (see
Appendix B).

For fixing $\Delta_a/g=-1/2$ of the blue sideband, we plot the $U_a/g$ dependence of $g_{aa}^{(2)}(0)$ and $n^{(a)}_s$ as shown in Figs.~\ref{Fig4}(e) and \ref{Fig4}(f). Interestingly, the single-photon purity can be increased by more than two orders of magnitude for large value of $U_a/g$, corresponding to $g_{aa}^{(2)}(0)<10^{-4}$ with $U_a/g>7.8$. The mechanism for generation of strong PB is ascribed to the interplay of the Stark-shift enhanced energy-spectrum anharmonicity and quantum interference suppressed two-photon excitation. In addition, the steady-state photon number $n^{(a)}_s$ is monotonically growing with increasing $U_a/g$, as displayed in Fig.~\ref{Fig4}(f). The large increased $n^{(a)}_s$ can be readily understood in part by the additional effective cavity-driven field. Remarkably, we show that the strong PB with large $n^{(a)}_s$ occurs at a moderate Stark shift when $U_a/g\geq 2.0$, which is important for experimental feasibility of cavity QED. Therefore, our proposal can be used to generate the high-quality deterministic single-photon source, which satisfies the high single-photon purity and large cavity output, simultaneously.

\begin{figure}[htb]
\includegraphics[width=0.9\columnwidth]{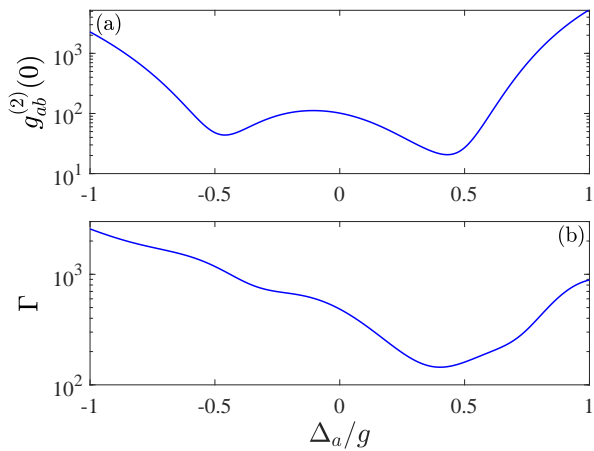}
\caption{$\Delta_a$ dependance of (a) cross-correlation function $g^{(2)}_{ab}(0)$ and (b) nonclassicality parameter $\Gamma$ based on quantum interference with $U_a/g=2$.} \label{Fig5}
\end{figure}

\section{Nonclassical correlated deterministic single-photon pairs}
\label{pairs}
To gain more insight into single-photon pairs, we study the nonclassical correlation for the generated photon pairs of the two-mode cavities. Remarkably, both the photon emissions in bimodal cavities could exhibit the strong PB by employing the Stark shift and quantum interference, which indicates that the nonclassical photon-photon correlations in the single-photon level are realized. In general, the pair of classical correlation functions for two-mode fields constrained by the well-known Cauchy-Schwarz inequality is given by~\cite{miranowicz2010testing}
\begin{align}
g^{(2)}_{ab}(0) \leq \sqrt{g^{(2)}_{aa}(0)g^{(2)}_{bb}(0)},
\end{align}
which gives rise to the bounded value of cross-correlation function $g^{(2)}_{ab}(0)$ in classical origin. A violation of Cauchy-Schwarz inequality can act as an unambiguous feature for measuring the nonclassicality of two-mode correlations~\cite{kimble2003,riedinger2016}. To proceed further, we introduce a dimensionless parameter for unraveling the nonclassicality, defined as
\begin{align}
\Gamma = \frac{g^{(2)}_{ab}(0)}{\sqrt{g^{(2)}_{aa}(0)g^{(2)}_{bb}(0)}} .
\end{align}
Note that $\Gamma>1$ represents the violation of Cauchy-Schwarz inequality for the nonclassical correlated deterministic single-photon pairs in our system.

Figure~\ref{Fig5}(a) shows the cross-correlation function $g^{(2)}_{ab}(0)$ as a function of $\Delta_{a}$ under the quantum interference. Obviously, $g^{(2)}_{ab}(0)$ exhibits a double-well structure near the single-photon resonance of red and blue sidebands. In the parameter regime of our numerical simulation, the values of $g^{(2)}_{ab}(0)$ are much larger than $1$, albeit the bimodal cavities photons exhibit the strong antibunching for both $g^{(2)}_{aa}(0)\ll 1$ and $g^{(2)}_{bb}(0)\ll 1$. In Fig.~\ref{Fig5}(b), we plot the nonclassicality parameter $\Gamma$ as a function of $\Delta_{a}$. The strong violation of the Cauchy-Schwarz inequality for single-photon pairs is observed, corresponding to $\Gamma \gg 1$, which exceeds the classical bound ($\Gamma=1$) clearly. More important, the bimodal cavities with the same dissipations possess the strong nonclassical correlations and equal photon emissions due to the deterministic parametric down-conversion process. Finally, the proposed system can be used as a high-quality two-mode single-photon source and single-photon pairs sources with hosting the strong nonclassicality, which could provide a broad physics community for applications in quantum information science~\cite{RevModPhys.87.347,flamini2018photonic,slussarenko2019photonic} and metrology~\cite{PhysRevX.1.021022,PhysRevLett.110.240402,RevModPhys.89.041003,RevModPhys.90.035006}.

\section{Conclusion}
\label{Con}
In conclusion, we propose an experimental scheme for realizing nonclassical correlated deterministic single-photon pairs by constructing cavity-enhanced deterministic parametric down-conversion in bimodal cavities. We show that the photon quantum statistics for the bimodal cavities exhibit the red-blue asymmetric feature by tuning the value of $U_{a}/g$, albeit the photon emissions are always the same. The strong photon blockade with $g_{aa}^{(2)}(0) < 10^{-4}$ is achieved by employing optical Stark-shift enhanced energy-spectrum anharmonicity and quantum interference suppressed two-photon excitation. In particular, the photon antibunching amplitude can be over two orders of magnitude smaller than the results in the absence of quantum interference. The advantages of our proposal is that it does not require strong atom-cavity coupling, which will facilitate the monitoring of quantum experimental feasibility in cavity QED. Interestingly, the observed deterministic single-photon pairs exhibit the strong nonclassical correlation with violating Cauchy-Schwarz inequality. Our study on realizing the deterministic photon sources can be extended to explore the multimode $N$-photon bundle states with strong nonclassicality natures~\cite{Deng21,munoz2014emitters,munoz2018,PhysRevLett.117.203602,PhysRevLett.127.073602}.

\section*{acknowledgment}\label{acknow}

This work was supported by the National Key R$\&$D Program of China (Grant No. 2018YFA0307500), NSFC (Grants No. 11874433 {\color{blue}and  No. 12135018}), and the Key-Area Research and Development Program of GuangDong Province under Grants No. 2019B030330001.

\section*{Appendix A: time-independent Hamiltonian}

In this appendix we present the details on the derivation of the effective Hamiltonian for the given level diagram and laser configurations in Fig.~\ref{Fig1}(a) of the main text. Under the rotating-wave approximation, the relevant Hamiltonian for a single atom trapped in bimodal cavities can be written as
\begin{align}\label{H1}
{\cal\hat{H}}_{1}/\hbar &= \omega_{a}\hat{a}^{\dagger}\hat{a} + \omega_{b}\hat{b}^{\dagger}\hat{b} + \omega_{e}|e\rangle\langle e| + \omega_{r}|r\rangle\langle r| + \omega_{r'}|r'\rangle\langle r'| \nonumber \\
&+ [g_{a}\hat{a}^{\dagger}|g\rangle\langle e| + g_{b}\hat{b}^{\dagger}|e\rangle\langle r| + g^{\prime}_{b}\hat{b}^{\dagger}|e\rangle\langle r'| + {\rm H.c.}] \nonumber \\
&+ [\Omega_{1}e^{i\omega_{1}t}|g\rangle\langle e| + \Omega_{2}e^{i\omega_{2}t}|e\rangle\langle r| \nonumber \\
&+ \Omega_{3}e^{i(\omega_{3}t+\theta)}|g\rangle\langle r'|  + {\rm H.c.}], \tag{A1}
\end{align}
where $\hat{a}^{\dagger}$ and $\hat{b}$ is the annihilation operator of cavity mode A and B), respectively, $\ket{m}$ is the electronically atomic excited state with $m =e, r, r'$, and $\hbar\omega_{m}$ is the bare transition energy of $\ket{g}\leftrightarrow\ket{m}$.

In the rotating frame, the unitary transformation is defined as
\begin{align}
{\cal U} &= \exp\{-i[\omega'_{a}(\hat{a}^{\dagger}\hat{a}+|e\rangle\langle e|) + \omega'_{b}\hat{b}^{\dagger}\hat{b}  \nonumber \\
&+ \omega_{3}(|r\rangle\langle r| + |r'\rangle\langle r'|)]t\},
\tag{A2}
\end{align}
with fixing $\omega'_{a}+\omega'_{b}=\omega_3$ and $\omega_3=\omega_{1}+\omega_{2}$. Then the internal-state Hamiltonian Eq.~(\ref{H1}) reduces to
\begin{align}\label{H2}
{\cal\hat{H}}_{2}/\hbar &= {\cal U}^{\dagger}{\cal \hat{H}}_{1}{\cal U} - i{\cal U}^{\dagger}\frac{\partial}{\partial{t}}{\cal U} \nonumber \\
&= \delta_{a}\hat{a}^{\dagger}\hat{a} + \delta_{b}\hat{b}^{\dagger}\hat{b} + \Delta_{1}|e\rangle\langle e| + \Delta_{r}|r\rangle\langle r| + \Delta_{3}|r'\rangle\langle r'| \nonumber \\
&+ [g_{a}\hat{a}^{\dagger}|g\rangle\langle e| + g_{b}\hat{b}^{\dagger}|e\rangle\langle r| + g^{\prime}_{b}(\hat{b}^{\dagger}|e\rangle\langle r'| + {\rm H.c.}]  \nonumber \\ &+ [\Omega_{1}e^{i(\Delta_{1}-\Delta_{2})t}|g\rangle\langle e| + \Omega_{2}[e^{-i(\Delta_{1}-\Delta_{2})t}|e\rangle\langle r| + {\rm H.c.}]  \nonumber \\ &+ [\Omega_{3}e^{i\theta}|g\rangle\langle r'| + {\rm H.c.}], \tag{A3}
\end{align}
where $\delta_{a} = \omega_{a}-\omega^{\prime}_{a}$, $\delta_{b} = \omega_{b}-\omega^{\prime}_{b}$, $\Delta_{1} =  \omega_{e}-\omega'_{a}$, $\Delta_{2} =  \omega_{e}-\omega_{1}$, and $\Delta_{3} =  \omega_{r'}-\omega_{3}$ are the single-photon detunings, and $\Delta_{r} = \omega_{r}-\omega_{1}-\omega_{2}$ is the two-photon detuning.

To proceed further, we rewrite the Hamiltonian (\ref{H2}) in the second quantized form
\begin{align}\label{H3}
{\cal\hat{H}}_{2}/\hbar &= \delta_{a}\hat{a}^{\dagger}\hat{a} + \delta_{b}\hat{b}^{\dagger}\hat{b} + \Delta_{1}\hat{\psi}_{e}^{\dagger}\hat{\psi}_{e} + \Delta_{r}\hat{\psi}_{r}^{\dagger}\hat{\psi}_{r} + \Delta_{3}\hat{\psi}_{r'}^{\dagger}\hat{\psi}_{r'} \nonumber \\
&+ [g_{a}\hat{a}^{\dagger}\hat{\psi}_{g}^{\dagger}\hat{\psi}_{e} + g_{b}\hat{b}^{\dagger}\hat{\psi}_{e}^{\dagger}\hat{\psi}_{r} + g^{\prime}_{b}\hat{b}^{\dagger}\hat{\psi}_{e}^{\dagger}\hat{\psi}_{r'} + {\rm H.c.}] \nonumber \\
&+[\Omega_{1}e^{i(\Delta_{1}-\Delta_{2})t}\hat{\psi}_{g}^{\dagger}\hat{\psi}_{e} +\Omega_{2}e^{-i(\Delta_{1}-\Delta_{2})t}\hat{\psi}_{e}^{\dagger}\hat{\psi}_{r} + {\rm H.c.}] \nonumber \\
& + [\Omega_{3}e^{i\theta}\hat{\psi}_{g}^{\dagger}\hat{\psi}_{r'} + {\rm H.c.}], \tag{A4}
\end{align}
where $\hat{\psi}_{g}$ and $\hat{\psi}_{m}$ denotes the annihilation field operator for the ground and excited state, respectively.

Taking into account the atomic spontaneous emissions, the Heisenberg equations of motion for the far-off resonance atomic field operators are given by
\begin{align}\label{eq of motion}
i\dot{\hat{\psi}}_{e} &= (\Delta_{1}-i\gamma_{e})\hat{\psi}_{e} + g_{a}\hat{a}\hat{\psi}_{g} + g_{b}\hat{b}^{\dagger}\hat{\psi}_{r} + g_{b}^{\prime}\hat{b}^{\dagger}\hat{\psi}_{r'} \nonumber \\
&+ \Omega_{1}e^{-i(\Delta_{1}-\Delta_{2})t}\hat{\psi}_{g} + \Omega_{2}e^{-i(\Delta_{1}-\Delta_{2})t}\hat{\psi}_{r} \nonumber \\
i\dot{\hat{\psi}}_{r'} &= (\Delta_{3}-i\gamma_{r'})\hat{\psi}_{r'} + g_{b}^{\prime}\hat{b}\hat{\psi}_{e} + \Omega_{3}e^{-i\theta}\hat{\psi}_{g}, \tag{A5}
\end{align}
where $\gamma_e$ and $\gamma_{r'}$ are the atomic spontaneous emission rate for $|e\rangle$ and $|r'\rangle$ states. Under the conditions $|\Delta_i|\gg \{g_a,g_b,\Omega_{i},\gamma_{e},\gamma_{r'}\}$ with $i=1,2,3$, the excited states $\ket{e}$ and $\ket{r'}$ can be adiabatically eliminated to yield
\begin{align}\label{adiabatical elimination}
\hat{\psi}_{e} &= -\frac{1}{\Delta_{1}-i\gamma_{e}}\left[g_{a}\hat{a}\hat{\psi}_{g} + \Omega_{1}e^{-i(\Delta_{1}-\Delta_{2})t}\hat{\psi}_{g} + g_{b}\hat{b}^{\dagger}\hat{\psi}_{r} \right. \nonumber \\
&\left.+ \Omega_{2}e^{-i(\Delta_{1}-\Delta_{2})t}\hat{\psi}_{r}\right], \nonumber \\
&\approx -\frac{1}{\Delta_{1}}\left[g_{a}\hat{a}\hat{\psi}_{g} + \Omega_{1}e^{-i(\Delta_{1}-\Delta_{2})t}\hat{\psi}_{g} + g_{b}\hat{b}^{\dagger}\hat{\psi}_{r} \right. \nonumber \\
&\left.+ \Omega_{2}e^{-i(\Delta_{1}-\Delta_{2})t}\hat{\psi}_{r}\right], \nonumber \\
\hat{\psi}_{r'} &= \frac{1}{(\Delta_{1}-i\gamma_{e})(\Delta_{3}-i\gamma_{r'})}\left[g_{a}g^{\prime}_{b}\hat{a}\hat{b}\hat{\psi}_{g} + g_{b}g^{\prime}_{b}\hat{b}^{\dagger}\hat{b}\hat{\psi}_{r} \right.\nonumber \\
&\left. + g^{\prime}_{b}\Omega_{1}e^{-i(\Delta_{1}-\Delta_{2})t}\hat{b}\hat{\psi}_{g}  + g^{\prime}_{b}\Omega_{2}e^{-i(\Delta_{1}-\Delta_{2})t}\hat{b}\hat{\psi}_{r} \right.\nonumber \\
&\left. - \Omega_{3}(\Delta_{1}-i\gamma_{e})e^{-i\theta}\hat{\psi}_{g}  \right],\nonumber \\
&\approx -\frac{\Omega_{3}\Delta_{1}e^{-i\theta}\hat{\psi}_{g}}{\Delta_{3}},
\tag{A6}
\end{align}
here the higher-order terms ($\sim 1/\Delta_1^2$ and $\sim 1/(\Delta_1\Delta_3)$) are safely ignored.

Substituting Eq.~(\ref{adiabatical elimination}) into Eq.~(\ref{H3}), one can derive an effective Hamiltonian for single atom-cavity system
\begin{align}\label{H4}
{\cal\hat{H}}_{3}/\hbar &= \Delta_{a}\hat{a}^{\dagger}\hat{a} + \Delta_{b}\hat{b}^{\dagger}\hat{b} + (\Delta_{r} - U_{a}\hat{a}^{\dagger}\hat{a} + U_{b}\hat{b}^{\dagger}\hat{b})\hat{\psi}_{r}^{\dagger}\hat{\psi}_{r} \nonumber \\
&+ [(g\hat{a}^{\dagger}\hat{b}^{\dagger}+\Omega)\hat{\psi}_{g}^{\dagger}\hat{\psi}_{r} + \eta e^{-i\theta}\hat{a}^{\dagger}\hat{b}^{\dagger}\hat{\psi}_{g}^{\dagger}\hat{\psi}_{g} + {\rm H.c.}]\nonumber \\
&- \left[\frac{g_a}{\Delta_1}\left(\Omega_1\hat{\psi}_{g}^{\dagger}\hat{\psi}_{g}+ \Omega_2\hat{\psi}_{g}^{\dagger}\hat{\psi}_{r} \right)\hat{a}^{\dagger} e^{-i(\Delta_{1}-\Delta_{2})t} + {\rm H.c.}\right]
\nonumber \\
&- \left[\frac{g_b}{\Delta_1}\left(\Omega_2\hat{\psi}_{r}^{\dagger}\hat{\psi}_{r}+ \Omega_1\hat{\psi}_{g}\hat{\psi}_{r} \right)\hat{b}^{\dagger} e^{-i(\Delta_{1}-\Delta_{2})t} + {\rm H.c.}\right],
\tag{A7}
\end{align}
where $\Delta_a = \delta_{a} - g^{2}_{a}/\Delta_{1}$ ($\Delta_b = \delta_b$) is the effective cavity-light detuning for cavity mode A (B), $g=-g_{a}g_{b}/\Delta_{1}$ ($\Omega=-\Omega_{1}\Omega_{2}/\Delta_{1}$) is cavity (pump) field induced two-photon Rabi frequency, $\eta = \Omega_{3}g_{a}g^{\prime}_{b}/\Delta_{1}\Delta_{3}$ is the effective cavity-driven amplitude, and $U_a=-g^{2}_{a}/\Delta_{1}$ ($U_b=-g^{2}_{b}/\Delta_{1}$) is the optical Stark shift for cavity mode A (B). We emphasize that the terms with the high-frequency prefactor $e^{\pm i(\Delta_1-\Delta_2)t}$ in Eq.~(\ref{H4}) are safely neglected when  $|(\Delta_1-\Delta_2)|\gg \{g_a,g_b,\Omega_{i=1,2}\}$. This condition can be easily satisfied by taking $\Delta_1\neq\Delta_2$, e.g., $\Delta_1=-\Delta_2$. Finally, we obtain the effective time-independent Hamiltonian (\ref{Hamiltonian2}) in the main text.

\begin{figure}[htb]
\includegraphics[width=0.99\columnwidth]{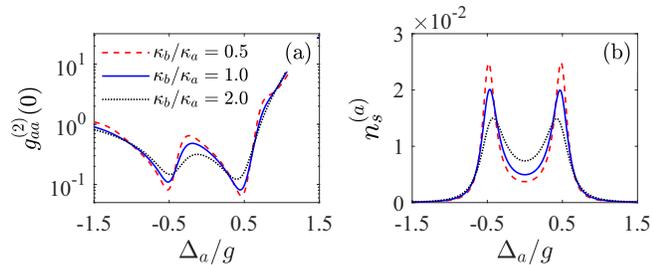}
\caption{$\Delta_a$ dependence of the second-order correlation function $g_{aa}^{(2)}(0)$ in (a) and steady-state photon number $n_s^{(a)}$ in (b) for different values of $\kappa_b$. Here the other parameters are $\eta/g=0$, $\kappa_a/\kappa=1$, and $U_a/g=2$.} \label{FigS2}
\end{figure}

\begin{figure}[htb]
\includegraphics[width=0.99\columnwidth]{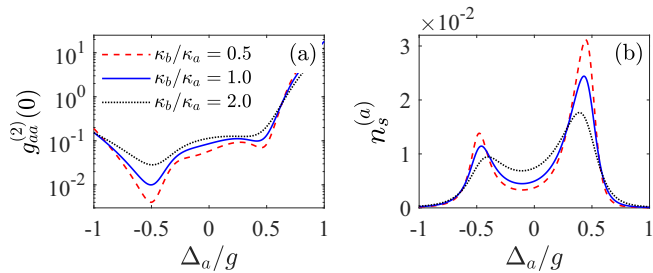}
\caption{$\Delta_a$ dependence of the second-order correlation function $g_{aa}^{(2)}(0)$ in (a) and steady-state photon number $n_s^{(a)}$ in (b) for different values of $\kappa_b$ using quantum interference conditions of Eq.~(\ref{QI}). Here the other parameters are $\kappa_a/\kappa=1$ and $U_a/g=2$.} \label{FigS4}
\end{figure}

\section*{Appendix B: The results with different bimodal cavities decays}

We present the numerical results of the photon quantum statistics with different bimodal cavities decays. In Figs.~\ref{FigS2}(a) and \ref{FigS2}(b), we plot the second-order correlation function $g_{aa}^{(2)}(0)$ and steady-state photon number $n_s^{(a)}$ for different values of cavity decay $\kappa_b$ with $\eta/g=0$. It is clear that the values of $g_{aa}^{(2)}(0)$ at both red and blue sidebands $\Delta_a/g=\pm1/2$ are monotonically increasing with decreasing the ratio $\kappa_b/\kappa_a$. Meanwhile, the steady-state photon number $n_s^{(a)}$ with respect to $\Delta_a$ remains exhibiting a red-blue symmetric profile. As can be seen, the peak values of $n_s^{(a)}$ increases as $\kappa_b/\kappa_a$ increases.

To proceed further, we investigate the effect of the decay rate $\kappa_b$ on the photon emissions in the presence of the quantum interference conditions in Eq.~(\ref{QI}). Figures~\ref{FigS4}(a) and \ref{FigS4}(b) shows $g_{aa}^{(2)}(0)$ and $n_s^{(a)}$ as a function of $\Delta_a$ for different values of cavity decay $\kappa_b$ by using the quantum interference conditions (\ref{QI}), respectively. It is cleat that the antibunching amplitude of $g_{aa}^{(2)}(0)$ at the blue sideband $\Delta_a/g=-1/2$ is rapidly increasing with decreasing the ratio $\kappa_b/\kappa_a$, albeit $n_s^{(a)}$ is slightly increased. As to the red sideband $\Delta_a/g=1/2$, we find that $n^{(a)}_s$ is obviously growing with decreasing $\kappa_b/\kappa_a$. Finally, we should remark that the qualitative results (see Fig.~\ref{FigS2} and Fig.~\ref{FigS4}) for both $g_{aa}^{(2)}(0)$ and $n_s^{(a)}$ are robust against the variations of the parameters $\kappa_a$ and $\kappa_b$.

%
\end{document}